# A SURVEY OF THE NORTHERN SKY FOR TeV POINT SOURCES

K. Wang,[1,a] R. Atkins,[2] W. Benbow,[3] D. Berley,[4] M.-L. Chen,[4,b] D. G. Coyne,[3] B. L. Dingus,[2] D. E. Dorfan,[3] R. W. Ellsworth,[5] A. Falcone,[6] L. Fleysher,[7] R. Fleysher,[7] G. Gisler,[8] J. A. Goodman,[4] T. J. Haines,[8] C. M. Hoffman,[8] S. Hugenberger,[9] L. A. Kelley,[3] I. Leonor,[9] M. McConnell,[6] J. F. McCullough,[3] J. E. McEnery,[2] R. S. Miller,[8,c] A. I. Mincer,[7] M. F. Morales,[3] P. Nemethy,[7] J. M. Ryan,[6] F. W. Samuelson,[8] B. Shen,[1] A. Shoup,[9] G. Sinnis,[8] A. J. Smith,[4] G. W. Sullivan,[4] O. T. Tumer,[1] M. O. Wascko,[1,d] S. Westerhoff,[3,e] D. A. Williams,[3] T. Yang,[3] G. B. Yodh[9] (The Milagro Collaboration)

[1]University of California, Riverside, CA 92521

[2]University of Wisconsin, Madison, WI 53706

[3]University of California, Santa Cruz, CA 95064

[4]University of Maryland, College Park, MD 20742

[5]George Mason University, Fairfax, VA22030

[6]University of New Hampshire, Durham, NH 03824-3525

[7]New York University, New York, NY 10003

[8]Los Alamos National Laboratory, Los Alamos, NM 87545

[9]University of California, Irvine, CA 92717

[a]Now at Fresher Information Corp., San Francisco, CA 94105

[b]Now at Brookhaven National Laboratory, Upton, NY 11973

[c]Now at University of New Hampshire, Durham, NH 03824-3525

[d]Now at Louisiana State University, Baton Rouge, LA 70803

[e]Now at Columbia University, New York, NY 10027

Abstract

A search for steady TeV point sources anywhere in the northern sky has been made with data from the Milagrito air-shower-particle detector. Over $3 \times 10^9$ events collected from 1997 February to 1998 May have been used in this study. No statistically significant excess above the background from the isotropic flux of cosmic rays was found for any direction of the sky with declination between -5° and 71.7°. Upper limits are derived for the photon flux above 1 TeV from any steady point source in the northern sky.

*Subject heading:* Gamma-rays: observations

1. INTRODUCTION

Many objects in the universe emit electromagnetic radiation via thermal processes. High-energy gamma rays can only be produced in non-thermal, energetic acceleration processes or by more exotic phenomena such as evaporating primordial black holes, topological defects, and neutralino annihilation. Observations of high-energy gamma rays have begun to reveal a picture of high-energy acceleration processes in the universe. Several reviews of high-energy gamma-ray astronomy have recently appeared (Ong 1998; Hoffman et al. 1999; Weekes 2000).

Cosmic gamma rays in the GeV regime can be detected directly with satellite-based detectors. The EGRET instrument (Thompson et al. 1993) surveyed the sky above 100 MeV and detected diffuse Galactic and extragalactic emission, five pulsars, at least 66 blazars (flat spectrum radio quasars and BL Lacertae objects), and about 170 "unidentified" objects found along the Galactic plane (Thompson et al. 1995; Hartman et al. 1999). Many of the blazars exhibit time variability, while the pulsars seem to be steady with no pulsed emission detected. The unidentified Galactic sources appear to be of two classes: relatively faint mid-latitude sources that may be associated with the Gould belt, and brighter low-latitude sources that may be Galactic pulsars or supernova remnants (Gehrels et al. 2000). A catalog of sources above 1 GeV contains 57 sources, including 10 that were not identified in the lower-energy catalogs (Lamb & Macomb 1997). This catalog includes the five pulsars and 21 blazars seen at lower energies, as well as 30 unidentified sources near the Galactic plane. The sensitivity of EGRET decreases at higher energies making variability difficult to detect.

Gamma-ray fluxes fall rapidly with increasing energy and the small size of the satellite-based detectors limits the maximum energy at which sources can be detected. Observations with EGRET extended up to ~100 GeV, while GLAST (Gehrels & Michelson 1999) aims to detect bright sources up to ~300 GeV. Ground-based detectors have much larger collection areas, and have detected sources above 250 GeV. A high-energy gamma ray interacts high in the earth's atmosphere producing an extensive air shower (EAS). Ground-based gamma-ray telescopes detect the products of EAS that survive to ground level.

Imaging atmospheric Cherenkov telescopes (ACT) have been successfully employed over the last decade in the energy region from ~250 GeV to 50 TeV. ACTs detect Cherenkov light produced in the atmosphere by the ultra-relativistic charged particles in the EAS. ACTs have detected emission from a number of sources including three pulsar-driven nebulae (the Crab, Vela, and PSR1706-44), and three blazars (Markarian 421, Markarian 501, and 1ES2344+514). The TeV emission from the pulsar-driven nebulae appears to be steady, while the blazars exhibit strong flares. However, because of their small field of view (~10 msr) and low duty factor (typically 10%), ACTs are not well suited to scan the sky for TeV sources. In fact, only a small fraction of the sky has been examined with imaging ACTs. A survey of the northern sky using a non-imaging ACT has been reported (Helmken, Horine & Weekes 1979; Weekes, Helmken & L'Heureux 1979). Ong (1998), Hoffman et al. (1999), and Weekes (2000) contain reviews of the ACT technique and observations.

Extensive air-shower particle detector arrays (EAS arrays) have been widely used to search for gamma-ray sources above ~50 TeV. An EAS array has typically consisted of a sparse array of scintillation counters that detects a small fraction of the charged EAS particles that reach ground level. An EAS array can operate 24 hours per day regardless of weather, and can simultaneously observe the entire overhead sky; an EAS array is able to observe every source in its field of view every day of the year. No convincing evidence for steady gamma-ray emission above 50 TeV has been obtained with these detectors (Alexandreas et al. 1991; McKay et al. 1993).

We report here on the results of a systematic search of the northern sky for steady point sources of TeV emission using data from the Milagrito detector, an EAS array sensitive to gamma rays in the 1-TeV region.

2. THE MILAGRITO DETECTOR

The Milagrito detector (Atkins et al. 2000), which operated from 1997 February to 1998 May, was the first EAS array to use a large volume of water as the detection medium. Milagrito used photomultiplier tubes (PMTs) deployed under water to detect the Cherenkov radiation produced in the water by relativistic charged EAS particles. Most of the shower particles (including the gamma rays, which produce electrons and positrons via pair production) falling within the area of the water can be detected with this technique. Milagrito had good sensitivity to primary gamma rays around 1 TeV, while maintaining the large-aperture and high-duty-factor attributes of an EAS array. Milagrito had a rather broad energy response with no well-defined energy threshold. The median energy for detected photons is ~3 TeV for a source at declination 40°, rising to ~7 TeV for a source at declinations of 20° and 60°.

Milagrito utilized 228 8"-diameter PMTs located in a plane on a 2.8 m x 2.8 m grid near the bottom of a large, covered pond of water. Data were taken with 1, 1.5, and 2 m of water above the PMTs. The pond is located at an altitude of 2650 m above sea level (atmospheric overburden = 750 g/cm$^2$) in the Jemez mountains west of Los Alamos, New Mexico (35.9° N, 106.7° W). As in a conventional EAS array, the direction of the primary gamma ray is reconstructed by measuring the relative times at which the individual PMTs were struck by light produced by particles in the shower front. To reduce the impact of stray light, only tubes registering pulses larger than 2 photoelectrons are used in the shower fits.

The angular resolution for cosmic-ray events was studied by comparing the incident directions separately reconstructed with two independent interleaved portions of the detector. The resulting angular resolution is in good agreement with expectations from a detailed Monte Carlo simulation (Atkins et al. 2000). However, this comparison is not sensitive to certain systematic effects such as those due to errors in the reconstructed shower core location. Observations of the cosmic ray shadow of the moon (Wascko et al. 1999) indicate that the simulation reproduces the actual detector point spread function quite well. These observations are for events initiated by cosmic rays, and it is the point spread function for events initiated by gamma rays that must be known to optimize the search for a point source of gamma rays. The simulations indicate that the angular resolution is 10-20% better for gamma-ray events. This is in general agreement with the shape of the signal from Markarian 501 observed with Milagrito (Atkins et al. 1999). The technique used to search for point-source emission, described below, depends only weakly on the detailed shape of the detector point spread function. Further details on Milagrito are available in Atkins et al. (1999).

3. ALL-SKY SURVEY TECHNIQUE

Milagrito had no ability to distinguish showers initiated by gamma rays from background showers initiated by charged cosmic rays. Because of the large number of isotropic background cosmic-ray showers, a strong point gamma-ray source would appear as an excess number of showers coming from a small region of the sky consistent with the Milagrito point spread function. The point spread function has wider tails than a Gaussian distribution, and narrows rapidly with increasing value of $N_{Fit}$, where $N_{Fit}$ is the number of PMTs participating in the fit to the shower plane. This analysis requires $N_{Fit} > 50$, a tighter cut than was used in Atkins et al. (1999) because this reduces the amount of data to be handled without significantly decreasing the sensitivity to point sources. For events with $N_{Fit} > 50$, the median value of the angular resolution is ~1°. This analysis uses 3.4 x $10^9$ events.

The method used involves dividing the sky into bins and examining each bin for an excess number of events. Rectangular sky bins rather than circular bins are used because they can be easily configured to cover the entire search region. With the $N_{Fit} > 50$ cut, the optimal bin on the sky in which to search for a point source is ~2° x 2°. The actual bin size in celestial coordinates is 2.22° [declination, $\delta$] x 2.22°/cos($\delta$) [right ascension, $\alpha$] with one abnormally-sized bin (at each $\delta$) ending at $\alpha = 360°$. Because the expected significance for a source depends only weakly on small changes of the bin size (Atkins et al. 1999), a bin a little larger than optimal was chosen to reduce the number of bins in the survey. A map consisting of a grid of 4319 bins covers the sky from $\delta = -3.9°$ to $\delta = 71.7°$. An unknown source may be located near an edge of a bin. In this case, the significance of a source, expressed in terms of the number of standard deviations above background, would be ~60% of the significance of a source located at the center of the bin. Thus three additional maps were also examined: one with bins shifted by 1.11° in $\delta$, one with bins shifted by 1.11°/cos($\delta$) in $\alpha$, and one with bins shifted by these amounts in both $\delta$ and $\alpha$. With this procedure, the largest loss in significance due to the location of a source relative to the center of a bin is ~15%. The survey with four maps covers the region from $\delta = -5°$ to $\delta = 71.7°$.

An accurate estimate of the expected background is needed in each bin. The expected background in a given bin depends upon its exposure and on the detector efficiency, which varies with the local coordinates ($\delta$, and hour angle, HA). The background estimation in each bin is determined using the data itself (Alexandreas et al. 1991, 1993). Random "background" events are generated for each observed event by associating the event local coordinates ($\delta$, HA) with times selected randomly from all event times recorded over a 2-hour period. New values of celestial coordinates are then calculated for each background event. Because the selection of a new event time amounts to a rotation of the celestial sphere with respect to the earth, this method changes only the $\alpha$ of an event but not its $\delta$. Fifteen background events are generated for each actual event so that the statistical error for the estimated background is small compared to the fluctuations in the signal map. The background events are then binned in the same manner as the data events: the expected background level is the number of background events falling into a source bin, divided by 15. It has been shown that this method produces background events that track changes in the trigger rate and are free of systematic errors (Alexandreas et al. 1991, 1993). Once signal and background maps have been produced, they are compared on a bin-by-bin basis and the statistical significance of the observed number of events being due to a fluctuation of the expected background is calculated using the method of Li & Ma (1983).

4. RESULTS

Figure 1a shows the distribution of the statistical significance of the observed number of events above the expected cosmic-ray background for each of the 4319 independent bins of the first map. None of these bins shows a significant excess (or deficit), considering the number of independent bins examined. The distributions for the shifted maps are shown in Figures 1b-d. The bins in a given map are independent of each other but they are not independent of the bins in the other maps. A Monte Carlo study indicates that the effective number of independent bins searched for a large excess to appear in any bin of the four maps is $\approx 2.5 \times 4319 \approx 10,800$. The largest observed excess is $3.9\sigma$, corresponding to a post-trials probability of ~50% of being due to a fluctuation of the background. Thus no strong steady source of TeV emission is observed in the northern sky for the period from 1997 February to 1998 May.

The sensitivity of Milagrito has been evaluated using the Monte Carlo simulation (Atkins et al. 2000). This has been verified by comparing the observed rate of gamma rays from Markarian 501 from 1997 February-October (Atkins et al. 1999) with the flux measured by air Cherenkov telescopes (Samuelson et al. 1998; Aharonian et al. 1999), and by comparing the observed rate of cosmic-ray events with the expected event rate given the measured fluxes reported in Asakimori et al. (1998) and Wiebel-Sooth, Biermann & Meyer (1998): the rates in these comparisons agree to within 10% (Atkins et al. 1999). While the close agreement of these rates may be partially fortuitous, it can be used to limit the systematic error on the energy scale to less than 30%. A direct measurement of the energy scale using the deflection of the observed shadow of the moon in the geomagnetic field implies that the error in the energy scale is less than ±85% (Wascko et al., in preparation).

The upper limit on the number of gamma-ray events in a bin varies according to the number of observed events in that bin. Rather than give gamma-ray flux upper limits on a bin-by-bin basis, we give upper limits for a typical bin as a function of declination. The upper limit on the flux from a source, $\Phi_\gamma$, at a given declination expressed in terms of the upper limit on the number of gamma-ray events, $N_\gamma$, for the source bin at that declination is:

$$N_\gamma = \int \Phi_\gamma(E)\, A_{EFF}(E, \delta, HA(t))\, \varepsilon\, dE\, dt\ ,$$

where $A_{EFF}(E, \delta, HA)$ is the effective area of Milagrito as a function of gamma-ray energy and the local coordinates of the source position, $HA(t)$ is the hour angle of the source as a function of time, and $\varepsilon$ is the fraction of events from a source that fall within the source bin. The time integral is taken over the experiment observation time. Assuming that an integral source spectrum is $\propto E^{-1.5}$, the resulting typical 90% confidence-level upper limit on the integral gamma-ray flux above 1 TeV for any steady source as a function of declination is shown in Figure 2. For comparison, the integral gamma-ray flux above 1 TeV from the Crab, measured by the Whipple collaboration (Hillas et al. 1998), is also shown in Figure 2.

We have also searched for steady emission from the four known northern TeV sources, the Crab, Mrk 421, Mrk 501, and 1ES2344+514. In each case, the bin in any of the maps that is most nearly centered on the actual source position was used. The results are shown in Table 1. Because no bin is exactly centered on the source position and the bin used in this analysis is larger than the optimal bin, a small loss of sensitivity is to be expected. The observed excess for the Crab ($1.1\sigma$) is in good agreement with the expected signal size ($1.3\sigma$), considering the expected 7% loss of sensitivity. The excess from Mrk 501 ($2.8\sigma$) is less than reported in Atkins et al. (1999) ($3.7\sigma$). This is partially due to the small expected loss in sensitivity discussed above (an expected loss of ~12%), and partially due to statistical fluctuations in the number of events passing the tighter $N_{Fit}$ cut and in the signal size in the larger bin used here.

5. CONCLUSIONS AND OUTLOOK

A systematic survey of the northern sky ($-5° < \delta < 72°$) for point TeV sources has been made with data from the Milagrito air shower detector. No source with a steady flux larger than ~5 times the gamma-ray flux above 1 TeV from the Crab has been observed for the period from 1997 February to 1998 May. A paper describing a search of the northern sky for episodic emission on a variety of time scales using data from Milagrito is in preparation.

Milagrito was dismantled in 1998 to install the Milagro detector. Milagro began data taking in 1999 December. Milagro has ~4 times better sensitivity than Milagrito due to its larger size and the ability to distinguish photon-initiated showers from background showers using a second layer of PMTs located under 6m of water. In addition, each photomultiplier tube in Milagro has a reflecting conical baffle to increase its light-collection area and eliminate late signals due to horizontally-traveling light. The energy response of Milagro is similar to that of Milagrito. An array of 175 4.5-$m^2$ water detectors is being installed over an area of 10,000 $m^2$ surrounding the pond, which should improve the sensitivity of Milagro by a further factor of 2. One of the goals of Milagro is to sensitively survey the northern sky for TeV gamma-ray sources over a wide variety of time scales.

| Source | $\alpha$ | $\delta$ | $\Delta(\alpha)$ | $\Delta(\delta)$ | On-source events | Excess events | Significance |
|---|---|---|---|---|---|---|---|
| Crab | 83.6° | 22.0° | -0.02° | 0.33° | 9.15 x $10^5$ | 1050 | 1.1$\sigma$ |
| Mrk 421 | 166.1° | 38.2° | 0.51° | -0.13° | 1.13 x $10^6$ | -850 | -0.8$\sigma$ |
| Mrk 501 | 253.5° | 39.8° | 0.40° | 0.31° | 1.15 x $10^6$ | 2980 | 2.8$\sigma$ |
| 1ES2344+514 | 356.2° | 51.4° | 0.04° | -0.24° | 1.06.x $10^6$ | 227 | 0.2$\sigma$ |

Table 1: Observations with Milagrito of the four known northern TeV sources. $\Delta(\alpha)$ [$\Delta(\delta)$] is the angular distance from the center of the examined bin to the actual source location in right ascension [declination], "On-source events" is the number of observed events in the source bin, and "Excess events" is the difference between the number of observed events and the estimated number of background events in that bin.

ACKNOWLEDGMENTS

We acknowledge the contributions of the many people who helped construct Milagrito. We especially thank R. S. Delay and M. Schneider, who were indispensable in the construction, maintenance and operation of Milagrito. This work was supported in part by the National Science

Foundation (Grant Numbers PHY-9722617, -9901496, -0070927, -0070933, -0070968, -0096256), The U. S. Department of Energy Office of High Energy Physics, The U. S. Department of Energy Office of Nuclear Physics, Los Alamos National Laboratory, the University of California, the Institute of Geophysics and Planetary Physics, The Research Corporation, and the California Space Institute.

REFERENCES

Alexandreas, D. E., *et al.* 1991 Ap. J. 383, L53

Alexandreas, D. E. *et al*. 1993, Nucl. Instrum. Meth. Phys. Res. A 328, 570

Aharonian, F., *et al*. 1999, A&A, 342, 69

Asakimori, K. *et al*. 1998, Ap. J. 502 278

Atkins, R., *et al.* 1999, Ap. J. 525, L25

Atkins, R, *et al.* 2000, Nucl. Instrum. Meth. Phys. Res. 449A, 478

Gehrels, N., Macomb, D. J., Bertsch, D. L., Thompson, D. J. & Hartman, R. C. 2000, Nature, 404, 363

Gehrels, N. & Michelson, P. 1999, Astroparticle Phys. 11, 277

Hartman, R. C., *et al*. 1999, Ap. J. Suppl. Ser. 123, 79

Helmken, H. F., Horine, E., & Weekes, T. C. 1979, Proc. 16[th] Int. Cosmic-Ray Conf. (Kyoto), 1, 120

Hillas, A. M., *et al*. 1998, Ap J. 503, 744

Hoffman, C. M., Sinnis, C., Fleury, P. & Punch, M. 1999, Rev. Mod. Phys.71, 897

Lamb, R. C., & Macomb, D. J. 1997, Ap. J. 488, 872

Li, T. & Ma, Y. 1983, Ap. J. 272, 317

McKay, T. A., *et al.* 1993, Ap. J. 417, 742

Ong, R. 1998, Physics Reports, 305, 93

Samuelson, F. W. *et al*. 1998, ApJ, 501, L17

Thompson, D. J., *et al*. 1993, Ap. J. Suppl. Ser. 86, 629

Thompson, D. J., *et al*. 1995, Ap. J. Suppl. Ser. 101, 259

Wascko, M., *et al.*1999, Proc. 26th Int. Cosmic-Ray Conf. (Salt Lake City), **7**, 218

Weekes, T. C., Helmken, H. F., & L'Heureux, J. 1979, Proc. 16$^{th}$ Int. Cosmic-Ray Conf. (Kyoto), 1, 126

Weekes, T. C. 2000, Physica Scripta, T85, 195

Wiebel-Sooth, B., Biermann, P. L., & Meyer, H. 1998, A & A 330, 389

FIGURES

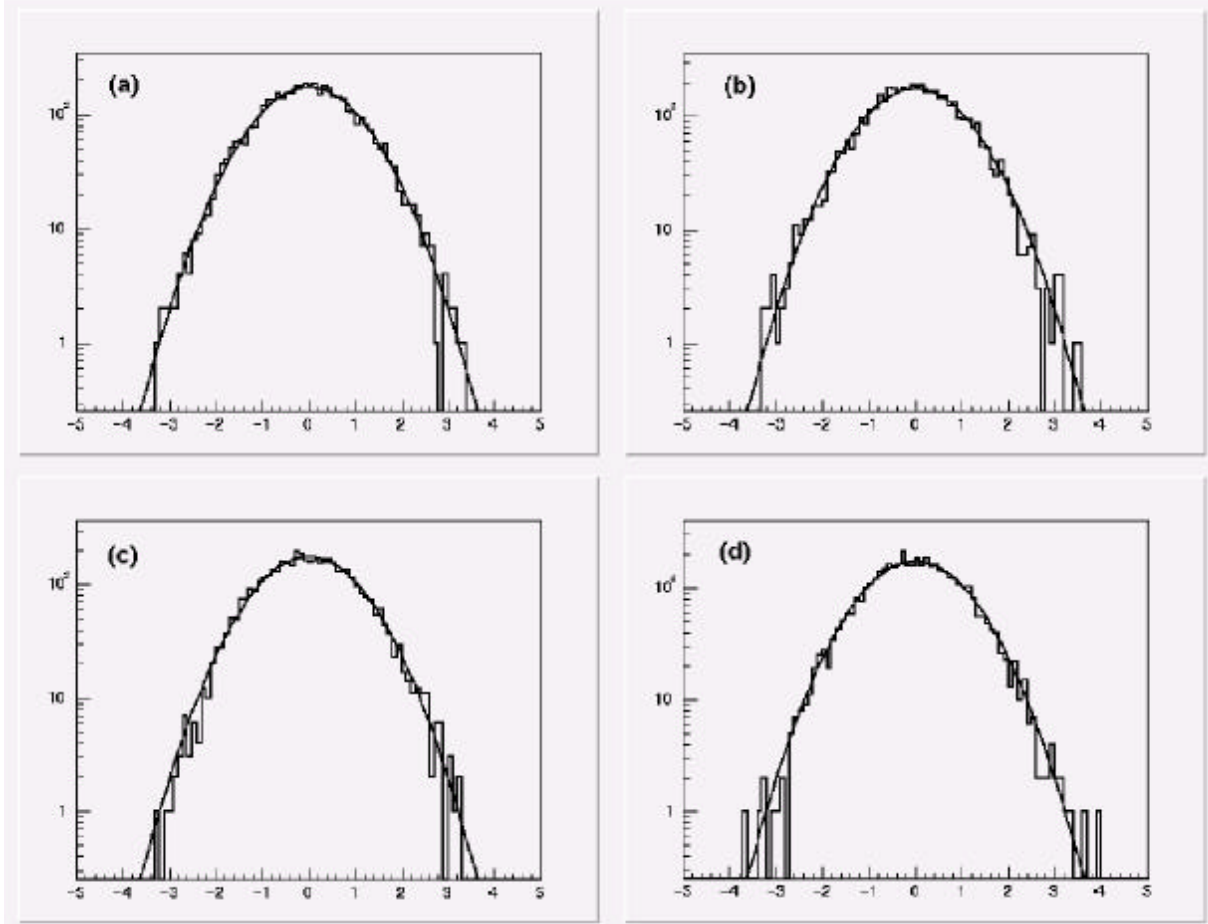

Fig. 1a. - The statistical significance for the observed number of events above the expected cosmic-ray background for each of the 4319 independent bins of the first sky map. The smooth curve is a Gaussian with unit width and area equal to 4319.

Fig. 1b-d. - The statistical significance for the bins of the three shifted maps.

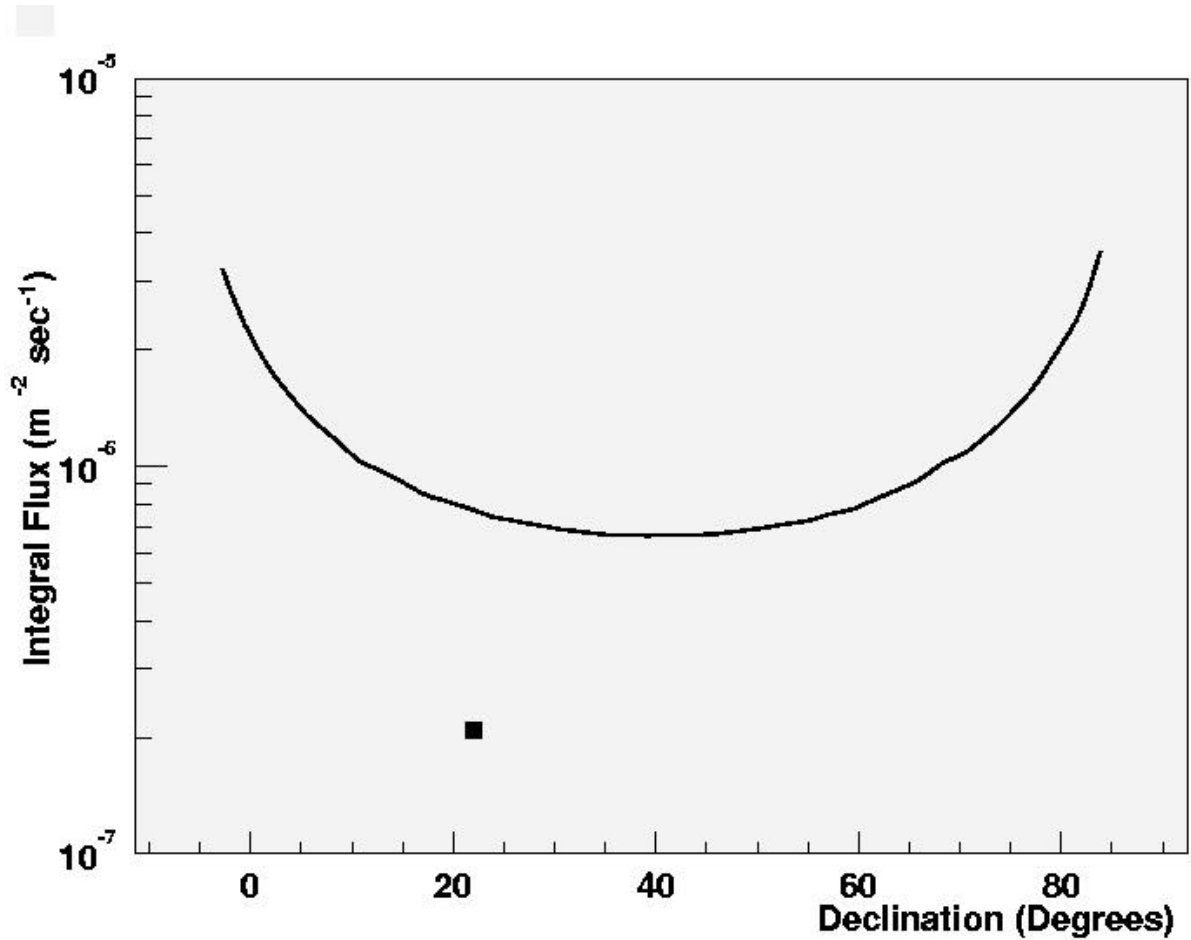

Fig. 2.- The typical 90% confidence level upper limit on the integral flux above 1 TeV for a steady source as a function of declination ($\delta$), assuming an integral flux $\propto E^{-1.5}$. The square shows the measured integral flux above 1 TeV from the Crab from Hillas (1998) for comparison.